# Thermal-Carrier-Escape Mitigation in a Quantum-Dot-In-Perovskite Intermediate Band Solar Cell via Bandgap Engineering

Ugur D. Menda[1], Guilherme Ribeiro[1], Jonas Deuermeier[1], Esther López[2], Daniela Nunes[1], Santanu Jana[1], Irene Artacho[2], Rodrigo Martins[1], Iván Mora-Seró[3], Manuel J. Mendes[1], and Iñigo Ramiro[1,2,*]

[1] CENIMAT/I3N, Departamento de Ciência dos Materiais, Faculdade de Ciências e Tecnologia, FCT. Universidade Nova de Lisboa and CEMOP/UNINOVA, 2829-516 Caparica, Portugal

[2] Instituto de Energía Solar - Universidad Politécnica de Madrid, ETSI Telecomunicación, Ciudad Universitaria, 28040 Madrid, Spain

[3] Institute of Advanced Materials (INAM), Universitat Jaume I (UJI), Castelló de la Plana 12071, Spain

*i.ramiro@upm.es

## Abstract

By harvesting a wider range of the solar spectrum, intermediate band solar cells (IBSCs) can achieve efficiencies 50% higher than conventional single-junction solar cells. For this, additional requirements are imposed to the light-absorbing semiconductor, which must contain a collection of in-gap levels, called intermediate band (IB), optically coupled to but thermally decoupled from the valence and conduction bands (VB and CB). Quantum-dot-in-perovskite (QDiP) solids, where inorganic quantum dots (QDs) are embedded in a halide perovskite matrix, have been recently suggested as a promising material platform for developing IBSCs. In this work, QDiP solids with excellent morphological and structural quality and strong absorption and emission related to the presence of in-gap QD levels are synthesized. With them, QDiP-based IBSCs are fabricated and, by means of temperature-dependent photocurrent measurements, it is shown that the IB is strongly thermally decoupled from the valence and conduction bands. The activation energy of the IB→CB thermal escape of electrons is measured to be 204 meV, resulting in the mitigation of this detrimental process even under room-temperature operation, thus fulfilling the first mandatory requisite to enable high-efficiency IBSCs.

# 1. Introduction

In the quest for highly-performant photovoltaic devices going beyond the Shockley and Queisser limit for single-gap solar cells, the intermediate band solar cell (IBSC) promises efficiencies as high as 63% at maximum solar concentration, and close to 50% under conventional one-sun illumination; in pair with triple-junction solar cells but with a simpler single-junction device structure.[1,2] The underlying idea is to reduce the optical losses by harvesting solar photons with less energy than the bandgap of the employed semiconductor. For this, a special kind of semiconductor, an intermediate band (IB) material, is required (**Figure 1**a). Such material exhibits a collection of in-gap levels that act as a third electronic band, the IB, additional to the valence and conduction bands (VB and CB). The IB must be optically coupled to the VB and the CB, so that electronic transitions can occur between each pair of bands through absorption or emission of photons. Thus, low-energy photons can contribute to increasing the photocurrent by promoting electrons from the VB to the CB via sequential two-photon absorption (TPA), using the IB as a steppingstone (see red and yellow transitions in **Figure 1**a). Even though the generation of additional photocurrent due to TPA has been demonstrated in different IB materials,[3–8] this process has been proven inefficient so far, mainly because of a too weak photon absorption for transitions involving the IB.[2,9,10]

The increase in photocurrent $J$ due to TPA will result in an increased solar cell efficiency, so long as the presence of the IB does not convey a significant degradation of the cell's operation voltage $V$.[1] In essence, this means that the ideal IB should not create additional non-radiative recombination paths that may hinder the collection of the photogenerated carriers. To that end, it is imperative that there is a null density of states (DOS) connecting the IB to the VB and the CB.[2,11] If a high DOS exists in between the IB and, for example, the CB (**Figure 1**c), the IB↔CB thermal excitation and relaxation of carriers will be too fast, effectively reducing the cell's voltage. This has been the case in the IBSCs investigated so far, in which cryogenic temperatures were required to mitigate the thermal coupling of the IB to the VB or the CB.[8,12–14] In contrast, if a clean band diagram is achieved, in which the IB is effectively isolated from the both the VB and the CB (**Figure 1**d), the energy split between bands is too high for thermal processes to take place even at room temperature (RT), leaving optical processes as the main coupling mechanism; first requisite for very-high-efficiency IBSCs.

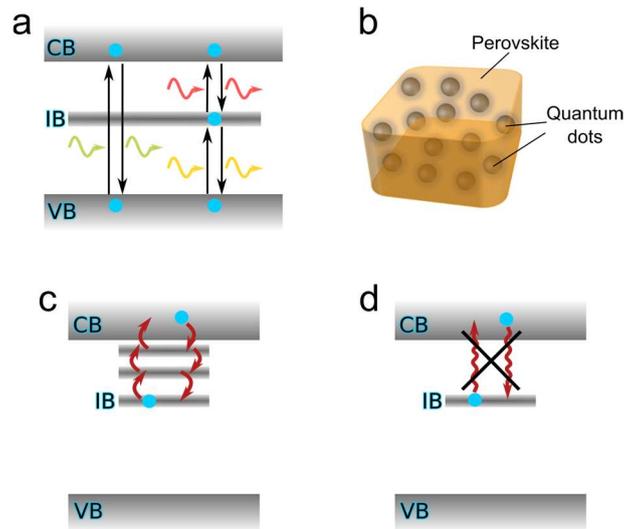

**Figure 1**. a) Schematic band diagram, depicting the photon absorption and emission processes in an ideal intermediate band (IB) material. b) Illustration of a quantum dot in perovskite (QDiP) material. c) Carrier thermal excitation and de-excitation is prominent when the IB is not effectively isolated from the CB. d) Non-radiative thermal processes are hindered when the IB and the CB are sufficiently split.

To date, epitaxial quantum dots (EQD) have been the preferred approach for engineering IB materials.[2,9] Grown within a semiconductor matrix, EQDs introduce confined electronic states inside the bandgap of the host semiconductor. Those confined electronic states form the IB. While the EQD technology has allowed proving the fundamental principles of IBSC operation,[15–17] it suffers from the two aforementioned shortcomings that have impeded fabricating efficient devices.[2] (i) EQDs are grown with low volumetric densities ($10^{15}$-$10^{16}$ dots/cm$^3$), which results in weak absorptivity for the transitions involving the IB. (ii) The size and shape of EQDs, resulting from the lattice-strain-driven Stranski-Krastanov growth method, produces a too large density of states in between the IB and the CB (or the VB), like in **Figure 1**c. While the low absorptance could be partially circumvented with optimized plasmonic[18] and/or photonic[19] light-trapping schemes, there is presently no effective solution to tackle the quantum-confinement issues of the EQDs geometry.

Recently, colloidal quantum dots (CQDs) embedded in a perovskite matrix (**Figure 1**b) were suggested as a promising IB material.[20] It has been argued that this technology can overcome the two main limitations found with EQDs.[2] (i) Films containing CQDs can be fabricated with volumetric densities three or four orders of magnitude higher than EQDs, which allows for strong absorbers; and (ii) the precise size control and the favorable aspect ratio of CQDs can be used to engineer IB materials with ideal band diagrams, like in **Figure 1**d. Recent computational simulations also point in this direction.[21]

Ning et al. first reported that CQDs can be embedded in a halide perovskite matrix giving rise to a composite material with high crystallinity and combined opto-electronic properties.[22] When dispersed in a perovskite solution, the CQDs act as nucleation centers in the perovskite crystallization process,[23,24] which can even result in improved crystallinity, film morphology and stabilization of a desired perovskite phase.[23,25–29] The appealing synergetic properties of quantum-dot-in-perovskite (QDiP) solids have quickly motivated intensive research in devices such as light-emitting diodes, photodetectors and solar cells.[30,31] In a pioneering work, Hosokawa et al. employed PbS quantum dots (QDs) in a methylammonium lead bromide perovskite matrix (PbS@MAPbBr$_3$) to fabricate a QDiP-based IBSC,[32] proving the production of photocurrent due to absorption in the embedded dots. However, no experimental evidence has been given yet in the line of demonstrating that the QDiP technology can actually overcome the fundamental shortcomings found so far in IBSCs.

In this work, we demonstrate a QDiP-based IBSC in which the thermal coupling of the IB to the conduction and valence bands is strongly mitigated at RT, thus fulfilling the first mandatory step for efficient IBSCs. This is achieved via QDiP bandgap engineering based on a judicious choice of QD and perovskite material. Furthermore, we show that our QDiP films have an absorption coefficient of almost $10^3$ cm$^{-1}$ in transitions involving the IB while keeping an excellent crystalline and morphological quality, proving that QDiPs can be strong multigap absorbers. These results experimentally verify the strong potential of QDiPs as a platform for developing high-efficiency IBSCs and other multigap devices.

## 2. Results and discussion

### 2.1 QDiP films

We decided to synthesize QDiP films based on PbS QDs and methylammonium lead iodide (MAPbI$_3$ or MAPI) perovskite, PbS@MAPI. The choice of this material combination will be justified later on. To fabricate the QDiP films, the QDs were dispersed in the perovskite solution prior to spin coating (see Methods). In a previous step, the original organic ligands of the PbS QDs, oleic acid (OA), were exchanged by iodide ligands by mixing the QDs in octane with a solution containing MAPI perovskite precursors (Supporting Figure S1), based on previously reported methods.[22,33] We will refer to the QDs prior and after this ligand-exchange process as OA-PbS QDs and MAPI-PbS QDs, respectively. Fourier transform infrared (FTIR) spectroscopy measurements confirmed the effective removal of the original organic ligands in the MAPI-PbS QDs (Supporting Figure S2). Coincidentally to previous reports,[27,34] the ligand-exchange process produces a red-shift in the excitonic absorption of the QDs, as shown in **Figure 2**a. The red-shift is reproducible and dependent on the QD exciton wavelength (Supporting Figure S3). We used this fact as a checkpoint to ensure homogenous conditions in all the samples used in our study.

We employed the MAPI-PbS QDs to fabricate QDiP films using different nominal QD concentrations $C_{QD}$, expressed in mg of QDs per ml of perovskite solution (see Methods). **Figure 2**b-c show exemplary planar scanning electron microscope (SEM) images of MAPI and QDiP films ($C_{QD}$ = 40 mg/ml). The QDiP film exhibits complete coverage, and a reduced grain size compared to the bare MAPI film. A grain size reduction in PbS@MAPI films has been reported for MAPI films with grain sizes similar to ours.[34] XRD spectra (**Figure 2**e-f) reveal high crystalline quality in QDiP films with dots of different diameters and $C_{QD}$ = 40 mg/ml, preserving the tetragonal crystal structure of the MAPI perovskite.[35] The characteristic XRD peaks of MAPI are only mildly decreased in the QDiP films, and there is no signature of the $PbI_2$ impurity. Furthermore, no signal related to PbS (see Supporting Figure S4) is detected, which means that the actual volumetric concentration of QDs in the QDiP film is still small. Nevertheless, the presence of the QDs is remarkably evident in the absorptance spectra (**Figure 2**d). With $C_{QD}$ = 40 mg/ml, the QDiP films typically absorb almost 3% of the light at the excitonic energy of the QDs, which, as we will show later on, represents the VB→IB transition in the IBSC framework. From the measured absorptance, we can calculate the absorption coefficient of the QDiP films (see Methods), which is around $7.5 \times 10^2$ cm$^{-1}$ at the QD excitonic energy (Supporting Figure S5). The fact that no trace of the PbS crystal is found in the XRD spectra, together with the measured absorption coefficient, point to the possibility of fabricating QDiPs with higher $C_{QD}$ and very strong absorption at energies above and below the perovskite bandgap, therefore suitable for highly-efficient IBSCs and other multigap optoelectronic devices.

Finally, photoluminescence (PL) measurements of the films (**Figure 2**g-h) reveal two important features. First, the emergence of the emission originating at the QDs in the QDiP (IB→VB emission in the IBSC framework) together with a strong quenching of the perovskite emission (CB→VB emission), suggesting efficient carrier transport from the MAPI host to the PbS QDs, in line with previous studies.[22,34] Secondly, the perovskite emission line, centered at 765 nm, is unaltered in the QDiP films. This means that the MAPI bandgap has not changed with the inclusion of the QDs

(also evidenced in **Figure 2**d), which in turn suggests that the presence of the QDs does not introduce significant strain in the overall lattice of the perovskite host. To sum up, all these results prove that our QDiP films have an excellent crystalline and morphological quality, as well as strong absorption and emission involving the in-gap QD states.

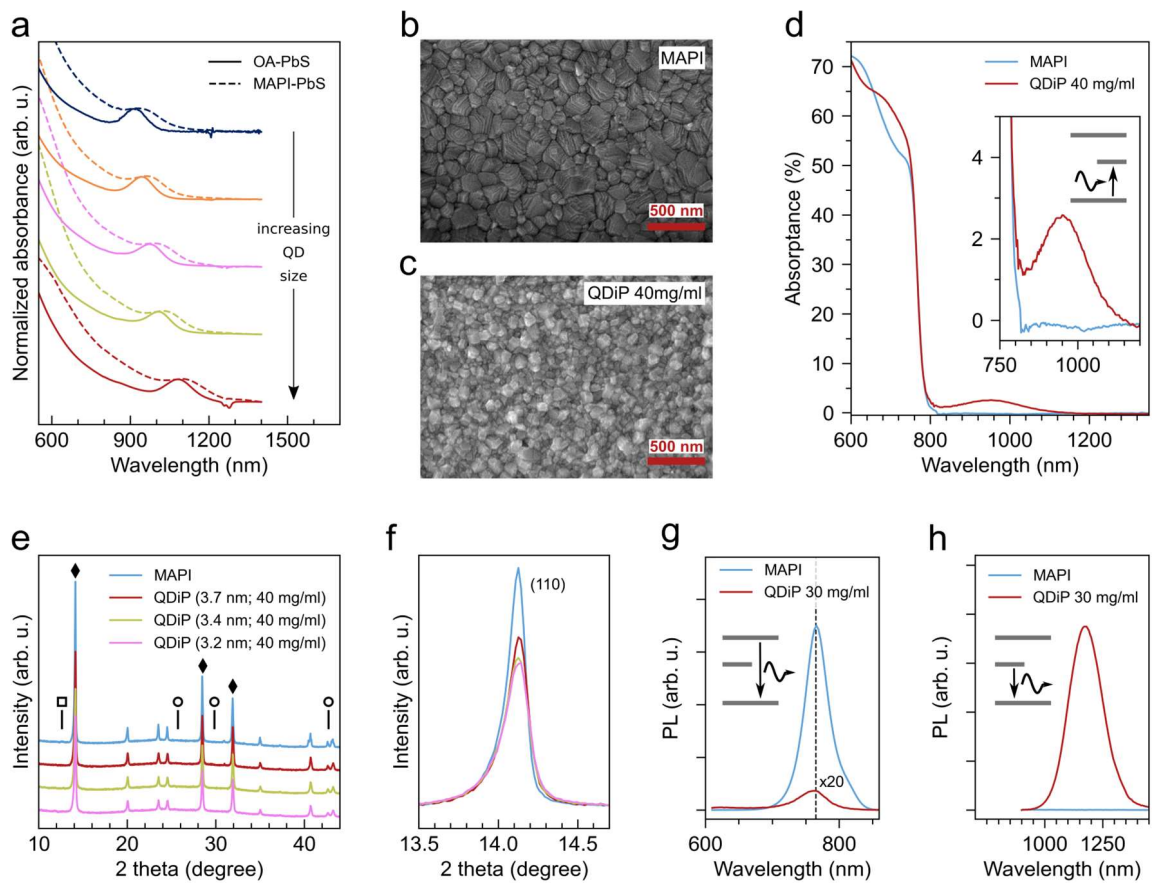

**Figure 2**. a) Normalized absorbance of OA-PbS QDs with different sizes and the corresponding MAPI-PbS QDs, dispersed in octane and DMF, respectively. b) and c) Planar SEM images of a MAPI film and a QDiP film (40 mg/ml). d) Exemplary absorptance spectrum of a MAPI film and a QDiP film (40 mg/ml). e) XRD spectra of MAPI and QDiP films with QDs of different sizes. The characteristic (110), (220) and (330) peaks of tetragonal MAPI, at 14.1°, 28.5°, and 31.9°, respectively, are indicated with filled diamonds. The open circles indicate the position of the characteristic peaks of PbS. The position of the characteristic peak of $PbI_2$ is indicted with an open square. f) Magnification of the XRD peaks at 14.1°. g) and h) Exemplary PL spectra of a MAPI film and a QDiP film (30 mg/ml). In g), the QDiP spectrum is multiplied by 20. A dashed line is plotted to indicate the central energy of MAPI emission at 765 nm.

The band positions in QDiPs is a topic that demands further exploration.[30] Regarding PbS@MAPI, theoretical works pointed to a type-I band alignment dependent on the QD size.[22,34] As explained before, an adequate band alignment is paramount for proper IBSC operation. Nonetheless, this aspect has not received enough attention in previous QDiP-IBSCs attempts.[32] In order to determine the band positions in our QDiPs, we have experimentally obtained the relevant energy levels of MAPI and the MAPI-PbS QDs (**Figure 3**a), and assumed a natural alignment in which the band offsets obtained from the individual components are preserved in the composite material. While some factors, such as interfacial lattice strain, can distort the band alignment in the heterocrystal, DFT calculations showed that keeping the natural offsets is a good approximation for QDiPs.[36]

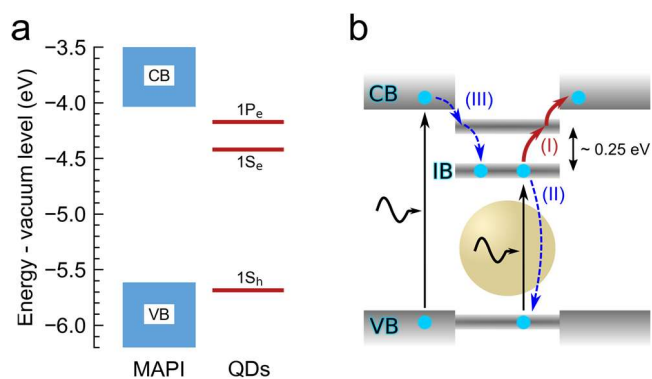

**Figure 3**. a) Experimentally determined energy levels of the MAPI and the MAPI-PbS QDs used to fabricate the QDiP-based IBSCs, except for $1P_e$, calculated from Equation 1 in Supporting Information. b) Resulting band diagram of the QDiP material used in our solar cells. Process I represents IB→CB thermal escape of electrons. Process II and III represent, respectively, IB→VB and CB→IB electron relaxation (of radiative or non-radiative nature).

The VB of MAPI and the ground states for holes, or $1S_h$, of the MAPI-PbS QDs were determined via ultraviolet photon spectroscopy (UPS) (Supporting Figures S6 and S7). The QDs used in these measurements had the same size (3.1 nm) than those used to build the solar cells studied later on. The QD size of the as-synthesized PbS QDs was determined from absorption measurements using a well-established formula.[37] The CB of MAPI and the ground states for electrons, or $1S_e$, of the QDs were obtained by addition of the measured optical bandgap. The second state for electrons, or $1P_e$, in the MAPI-PbS QDs is obtained using the empiric formula for the $1S_e$→$1P_e$ transition in PbS QDs with iodide ligands reported in the literature[38] (Equation 1 of the Supporting Information). The formula is validated for dots in the 5–8.5 nm range, so an extrapolation to our 3.1 nm dots was employed, obtaining 0.25 eV. We predict that the actual $1S_e$-$1P_e$ energy difference will be somewhat smaller in the composite material, given the lower potential barrier felt by the QDs in the QDiP as compared to the reference case of bare QD film.

The alignment of the energy levels of the MAPI and the MAPI-PbS QDs is the reason for our choice of materials. Indeed, when combined into a QDiP film, the PbS@MAPI will have a close-to-ideal band alignment from the point of view of an IB material (**Figure 3**b), i.e., an isolated IB in between the CB and the VB, leaving otherwise unperturbed the band diagram of the host semiconductor to avoid undesired potential barriers or extra recombination paths.[2]

**2.2 QDiP-based IBSCs**

We have fabricated IBSCs using QDiP ($C_{QD}$ = 40 mg/ml) as the absorbing material (QDiP SC), and control solar cells using bare MAPI (MAPI SC). The device structure is indicated in **Figure 4**b, where the only difference between the MAPI SC and the QDiP SC is the absorber layer. **Figure 4**a,c show exemplary cross-sectional SEM images of the solar cells. Both the MAPI and QDiP layers exhibit good morphology, forming flat and continuous films. The QDiP layer was constantly thicker than the MAPI counterpart, approximately 410 nm vs 360 nm. A detailed description of the different layers in the devices, as well as the device fabrication process, can be found in the Methods section.

External quantum efficiency (EQE) measurements (**Figure 4**d-e) allow assessing the IB-related photo-generation processes in the QDiP solar cells. In **Figure 4**d we can see that the response to photons absorbed in the perovskite host of the QDiP SC is reduced compared to the MAPI SC, which implies a reduced carrier collection. In **Figure 4**e we have plotted a zoom for wavelengths longer than the MAPI bandgap. The presence of the QDs in the QDiP SC produces a current response to low-energy (below-bandgap) photons, demonstrating the capability of the PbS@MAPI QDiP to extend the photon-energy range that can be harvested in a solar cell. Nevertheless, note that the EQE tracing is performed via a spectral sweep using monochromatic illumination at each wavelength. Therefore, this measurement does not allow the two-photon absorption (TPA) mechanism depicted in **Figure 1**a which would produce the desired IB-mediated photo-generation. As such, the ideal IB operation would be undetectable in the EQE spectra. This explains why the EQE signal in **Figure 4**e is weaker than one could directly expect when considering the pronounced below-bandgap absorptance of our QDiP films (**Figure 2**d). Indeed, QDiP films similar to those used in our QDiP SC absorb around 3% percent of the light at the QD excitonic wavelength. In contrast, the measured EQE is only around 0.2% in that energy range, which represents a greater than 10-fold reduction. However, rather than being a negative effect, this is a sign of proper IBSC operation owing to an adequately engineered band alignment, as we describe next.

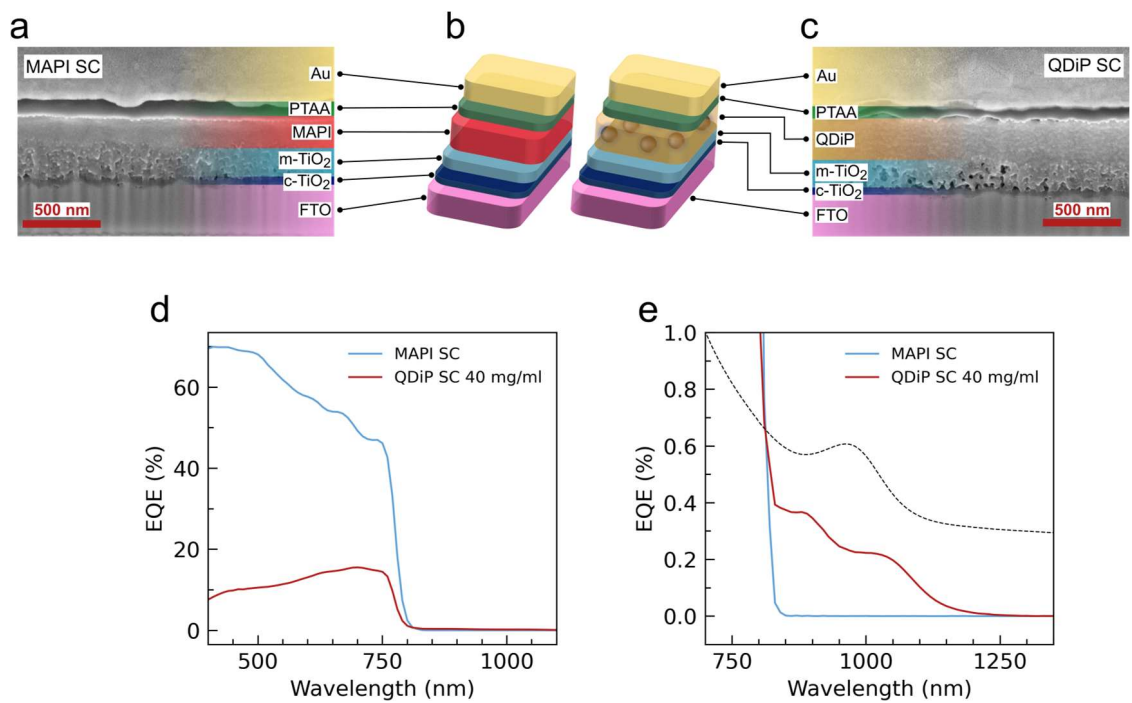

**Figure 4**. a) Cross-sectional SEM image of a reference MAPI SC. b) MAPI SC and QDiP SC layer structures. c) Cross-sectional SEM image of a QDiP SC. d) and e) EQE at room temperature (RT) of the MAPI SC and QDiP SC. In e), the dashed line plots the absorbance spectrum of the MAPI-PbS QDs solution used in the fabrication of the QDiP absorber layer.

We will use **Figure 3**b as graphical support to explain the EQE results. In our experiments, photons can be absorbed in the perovskite host of the QDiP, triggering a VB→CB transition, or in the QDs, triggering a VB→IB ($1S_h$→$1S_e$) transition. Since low-energy photons that could optically trigger the $1S_e$→$1P_e$ transition are not present in the experiment, electrons in the IB would require phonon assistance to escape from the IB into the CB (process I in the figure). However, in our QDiP films the $1S_e$-$1P_e$ energy difference is estimated to be around 250 meV, much higher that the lattice thermal energy at RT (≈ 26 meV). The IB→CB thermal carrier escape is therefore strongly hindered even at RT, which impedes the collection of carriers and results in low EQE values. Instead of being collected as photocurrent, most of the electrons in the IB relax back into the VB (process II). This is what is actually desired in a IBSC, where photocurrent involving absorption via the IB should only result from a TPA process. In fact, strong $1S_e \rightarrow 1P_e$ absorption has been reported in PbS QDs,[38] so one would expect to be able to measure photocurrent in our devices following TPA. First, an electron would be optically pumped from the VB to the IB, and then it would be promoted from the IB to the CB, following an optical $1S_e \rightarrow 1P_e$ excitation and subsequently being injected from the QD into the perovskite. Unfortunately, at the low energies of the $1S_e \rightarrow 1P_e$ transition in our QDiP films (0.25 eV or 5 μm), the glass/FTO substrate of the samples is completely opaque, so we could not measure any TPA-related photocurrent in our devices. The use of smaller PbS QDs with a larger $1S_e \rightarrow 1P_e$ transition would help in future studies to overcome this limitation.

The band diagram illustrated in **Figure 3**b also explains the reduction of the EQE at high photon energies. After being pumped into the CB, some of the electrons are captured by the QDs thereby relaxing to the IB (process III). As explained before, once in the IB, most of the electrons will relax back to the VB, and only a fraction will escape into the CB and contribute to the photocurrent. Hence, additional IB→CB optical pump would be required to recover the apparently degraded EQE at high photon energies.

To verify our explanation, we have performed temperature-dependent quantum efficiency measurements (EQE vs $T$). The idea is that, if the IB→CB carrier escape is thermally activated, the EQE at the wavelength range absorbed in the QDs should strongly diminish when the temperature is lowered. More importantly, if, as we claim, the carrier escape is mitigated even at RT, the same trend should exist for $T$ higher than RT. Additionally, we expect the EQE at short wavelengths (photons absorbed in the perovskite host of the QDiP) to be also dependent on $T$ since, as we discussed, also in this case the effectivity of carrier collection is affected by CB→IB electron relaxation into the QDs.

**Figure 5**a-b show the EQE vs $T$ spectra of the QDiP SC, in semilog and linear scale, respectively. For comparison, the EQE vs $T$ spectra of the MAPI SC are plotted in **Figure 5**c. The samples were measured in the 246 K – 306 K range. Higher temperatures were not explored to avoid the MAPI phase transition from tetragonal to cubic at around 310 K.[39] As expected, at wavelengths longer than the MAPI bandgap, the EQE of the QDiP SC is strongly dependent on temperature. At short-wavelengths, the EQE is also temperature dependent, although in weaker fashion. In contrast, the EQE of the MAPI SC only barely depends on temperature. To quantify the temperature dependence, we have extracted the activation energies $E_A$ of the different thermal processes from Arrhenius plots of the EQE, presented in **Figure 5**d. The plots are acquired for a selected short wavelength (700 nm) in both the MAPI and the QDiP samples, and for the approximative

wavelength of the excitonic response (1030 nm) in the QDiP SC. All three plots can be precisely fitted to the Arrhenius equation even for $T >$ RT. For the QDiP sample at 1030 nm, we obtain $E_A$=204 meV, in tune with the $1S_e$-$1P_e$ energy difference, estimated somewhat smaller than 250 meV. This high $E_A$ is responsible for the measured mitigation of the IB→CB thermal carrier escape.

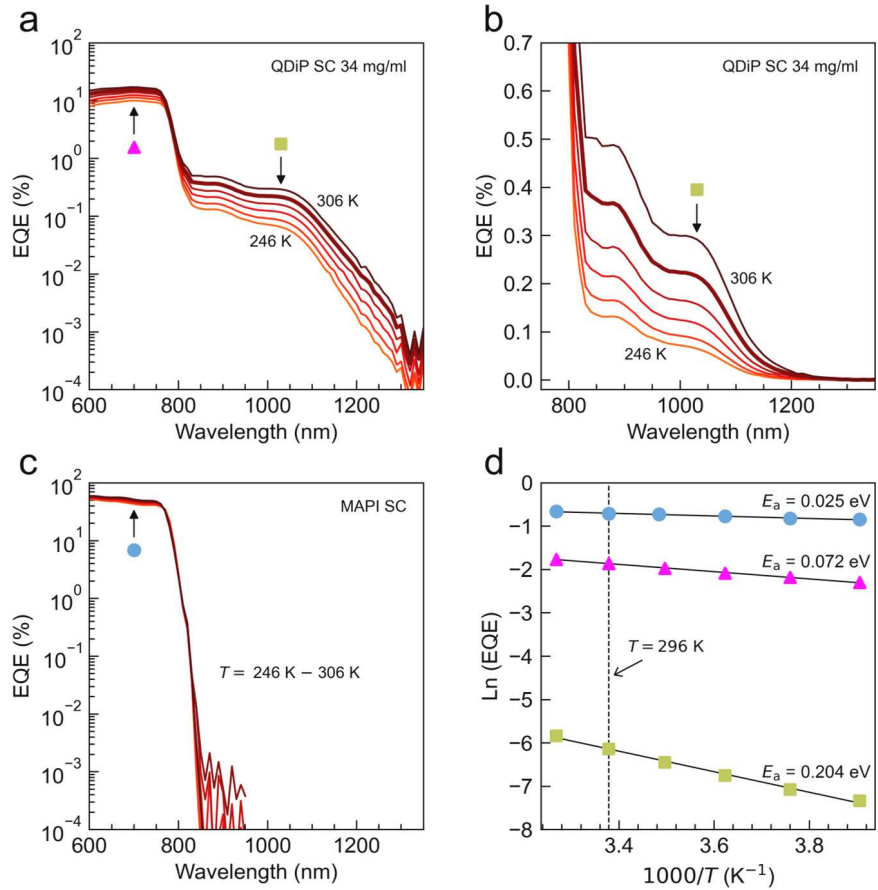

**Figure 5**. a,b) EQE vs $T$ of the QDiP SC in semilog (a) and linear (b) scale. The thicker line represents the room-temperature of our laboratory, $T$ = 296 K. c) EQE vs $T$ of the MAPI SC (semilog scale). d) Arrhenius plot of the EQE at 700 nm (triangles and circles) in the QDiP SC and MAPI SC, and at 1020 nm (squares) in the QDiP SC. The activation energies obtained for each case are indicated.

The activation energies obtained for the EQE at 700 nm are 25 meV and 72 meV, for the MAPI SC and the QDiP SC, respectively. Considering the MAPI SC, although weak, the temperature dependence implies that the collection efficiency of the photo-generated electrons and holes is less than one at RT. The observed trend is opposite to what would be expected from the transport properties of MAPI, as both mobility and diffusion length increase with lower temperature.[39] Such value of $E_A$ could be related to the exciton binding energy of MAPI, indicating that a fraction of the free electrons and holes relax into excitons and recombine before dissociation. In fact, the measured $E_A$ is comprised within the reported values of the exciton binding energy in MAPI,[40] ranging from a few meV to a few tens of meV. Another possibility for the weak temperature dependence of the EQE of the MAPI SC would be sub-optimal transport in the hole or electron transport layers.[41] On the other hand, the $E_A$ of the QDiP sample at 700 nm is almost 3 times as large, consistent with part of the photo-generated free electrons being trapped in the QDs (either in the $1S_e$ or higher-energy confined states), requiring thermal activation to be effectively collected.

## 3. Conclusion

The inclusion of colloidal PbS QDs in a MAPbI$_3$ matrix led to a composite material with excellent opto-electronic properties for the development of IBSCs, surpassing the potentiality showed by epitaxial QDs grown by the lattice-strain-driven Stranski-Krastanov method, thereby advancing a new avenue for the realization of IBSCs via a highly-scalable fabrication approach.

The confined electronic levels of the QDs form an effectively isolated IB, radiatively coupled to the electronic bands of the perovskite host, which is the first mandatory step for efficient IBSCs. The QDiP films showed strong absorption in the QDs while keeping fine structural and morphological quality. Moreover, we have determined a close-to-ideal band alignment between the MAPI-capped PbS QDs and the MAPI host, from the point of view of an IBSC. The significant IB-CB energy split, in the order of 250 meV, is responsible for the large activation energy of the IB→CB carrier thermal escape, making this detrimental process very inefficient at RT; a long-sought milestone in the IBSC community.

This work experimentally sets QDiPs as a very promising platform for the development of high-efficiency IBSCs, able to solve the main drawbacks encountered so far.

## 4. Methods

*PbS QD synthesis*. PbS QDs synthesis was adapted from previously reported multi-injection procedures. Briefly, a mixture of lead oxide (PbO), 1-octadecene (ODE), and oleic acid was degassed for two hours at 90 °C under vacuum. After degassing, the solution was placed under N$_2$ atmosphere, and a specific reaction temperature was set. A solution of hexamethyldisilathiane (TMS)$_2$S in ODE was quickly injected. After the injection, heating was stopped, and the solution was let to naturally cool down to room temperature. QDs were purified three times by precipitation with acetone and ethanol and redispersed in anhydrous octane.

*QD ligand exchange*. The as-synthesized OA-PbS QDs were diluted in octane with the desired concentration (mg/ml). OA-PbS QDs (in 2 ml of octane) were added into a low-concentration (0.25M) MAPI solution (in 2 ml of DMF) and stirred at room temperature until a clear transition of the QDs from octane to DMF was observed (Supporting Figure S1). Then, the octane with the OA ligands was removed and the QDs were washed with pure octane three times, to remove all the organic residue, and centrifuged with added toluene for the precipitation of dots. Finally, the supernatant was disposed, and the dots were dried under vacuum.

*MAPI and QDiP film fabrication*. The MAPI solution is prepared with 461 mg of $PbI_2$ and 159 mg of MAI in DMF and DMSO (v/v = 9:1), and stirring at 70 °C for 4 hours. QDiP solutions were prepared by dispersing the dried QDs after the ligand-exchange process in the MAPI solution. The concentration of the QDs in QDiP solutions is defined depending on the initial OA-PbS QD mass used in the ligand-exchange process, for consistency. For instance, if the ligand-exchange process starts using 40 mg of OA-PbS dots with oleic acid, assuming high-efficiency of the process and neglecting the contribution to the mass of the ligands, the concentration of MAPI-PbS in QDiP is determined as 40 mg/ml when used in 1 ml MAPI solution. MAPI and QDiP films were deposited on glass or FTO coated glasses by spin coating with the antisolvent washing method. The layers are deposited on glass for FTIR, XRD, UV-Vis spectrophotometry and PL measurements, and on FTO coated glass for SEM and UPS measurements. The deposition starts with a slower spin rate for 10 s, followed by 5000 rpm for 20s. Chlorobenzene was used as the antisolvent and dropped during the second step of the film growth. The films were annealed at 100 °C for 10 minutes to finalize the crystallization.

*Solar cell fabrication*. FTO coated glasses were cleaned by conventional methods, after which a UV-$O_3$ treatment was applied for 15 minutes to remove the organic compounds. The electron transport layer (ETL) is composed by a compact $TiO_2$ (c-$TiO_2$) and a mesoporous $TiO_2$ (m-$TiO_2$) layer. The c-$TiO_2$ film was spin coated at 4000 rpm for 30 s with a solution prepared with 180 μl TTIP, 18 μl HCl in 2.5 ml of ethanol. After the deposition, the layers are dried at 120 °C for 10 minutes and annealing at 500 °C for 30 minutes. For the m-TiO2 layer deposition, 150 mg of titania paste (Sigma Aldrich) was dissolved in 1 ml ethanol and spin-coated for 12 seconds at 4000 rpm, followed by a drying step at 100 °C for 5 minutes and sintering at 450 °C for 30 minutes in ambient air. The substrates were, subsequently, loaded into a glove box for the deposition of the absorber layer and the hole transport layer (HTL). The absorber layer (MAPI or QDiP) was deposited as previously described. As HTL, PTAA was dissolved in toluene (12 mg/ml), adding 10.5 µl of LiTFSI in acetonitrile (170 mg/ml) and 5.6 µl of 4-tert-butylpyridine. The solution was spin-coated at 1500 rpm for 30 s. The gold electrode was deposited by e-beam evaporation, forming rectangular devices of approximately 12 mm².

*Material characterization.* FTIR measurements were taken using a Thermo Nicolet 6700. Absorbance, transmittance, and reflectance measurements were taken using a Perkin Elmer Lambda 950. PL measurements were taken using a Horiba Fluorolog. XRD patterns were collected on a PANalytical XPert Pro X-ray powder diffractometer using Cu-Ka radiation. Planar SEM measurements were taken in a Hitachi Regulus 8220. Cross-section SEM observations were carried out using a Carl Zeiss AURIGA CrossBeam FIB-SEM workstation. The Ga+ ions were accelerated to 30 kV at 100 pA and the etching depth was kept around 1 µm. Ultraviolet photoelectron spectroscopy was done with a Kratos Axis Supra spectrometer. The source was a helium gas-discharge lamp set to He I emission (21.22 eV). The pass energy was set to 5 eV and the step size was 0.025 eV. The Fermi level of the spectrometer was calibrated for sputter-cleaned silver. The measurements were conducted in the dark with magnetic immersion lens turned off. The work function was determined by linear extrapolation of the secondary electron edge. The valence band maximum was extracted by linear extrapolation. In the case of the MAPI samples, a semi-logarithmic scale was used, since it has been shown that this method yields more realistic results.[42] The energy values obtained for PbS QDs were corrected following the work by Miller et al.[43]

*Solar cell characterization.* EQE measurements were carried out in a house-made setup using a QTH lamp as light source. The wavelength selection was made using a 0.25-m monochromator with appropriate order sorting filters. The current detection was performed using lock-in techniques, and calibrated Si and Ge detectors to characterize the incident spectral power. Temperature-dependent measurements were carried out in a closed-cycle He cryostat equipped with a heater. Prior to each measurement, the samples were kept during 25 min at a constant temperature to guarantee thermal stability.

*Calculation of $\alpha$.* Since the substrate is completely transparent to the wavelength range of interest, $\alpha$ is obtained, using the measured transmittance ($T$) and reflectance ($R$), directly from $A_{int} = 1 - e^{-\alpha}$, where $t$ is the film thickness, and $A_{int} = A/(1-R)$. $A = 1 - T - R$ is the measured absorptance. $A_{int}$ represents the intrinsic absorptance of the film, irrespective of the external medium, considering only the light that enters the film.

**Supporting Information and data availability**

The Supporting Information can be found at the end of this file. All raw data generated during the current study are available from the corresponding author upon reasonable request.

**Acknowledgements**

The authors acknowledge funding from the European Union's Horizon 2020 research and innovation programme under the Marie Sklodowska-Curie grant ENLIGHTEN (agreement No 891686) and SYNERGY (H2020-WIDESPREAD-2020-5, CSA, Grant No. 952169), as well as European Union's "NextGenerationEU"/PRTR. The work was also supported by national funds via FCT (*Fundação para a Ciência e Tecnologia,* I.P.) under the projects LA/P/0037/2020, UIDP/50025/2020 and UIDB/50025/2020 of the Associate Laboratory Institute of Nanostructures, Nanomodelling and Nanofabrication—i3N. I. Ramiro acknowledges funding through the Grant. No RYC2021-034610-I (Ramón y Cajal Fellowship), funded by MCIN/AEI/10.13039/501100011033 and the European Union «NextGenerationEU»/PRTR. G. Ribeiro also acknowledges funding from FCT, I.P. through the grant SFRH/BD/151095/2021.

Supporting Information

**Thermal-Carrier-Escape Mitigation in a Quantum-Dot-In-Perovskite Intermediate Band Solar Cell via Bandgap Engineering**

Ugur D. Menda[1], Guilherme Ribeiro[1], Jonas Deuermeier[1], Esther López[2], Daniela Nunes[1], Santanu Jana[1], Irene Artacho[2], Rodrigo Martins[1], Iván Mora-Seró[3], Manuel J. Mendes[1], and Iñigo Ramiro[1,2,*]

The second state for electrons, or 1P$_e$, in the MAPI-PbS QDs is obtained using Equation 1, reported for the 1S$_e$→1P$_e$ transition in PbS QDs with iodide ligands, where $d$ is the size (diameter) of the dots:

$$1P_e - 1S_e = \frac{1.57}{d} - \frac{2.5}{d^2} \qquad (1)$$

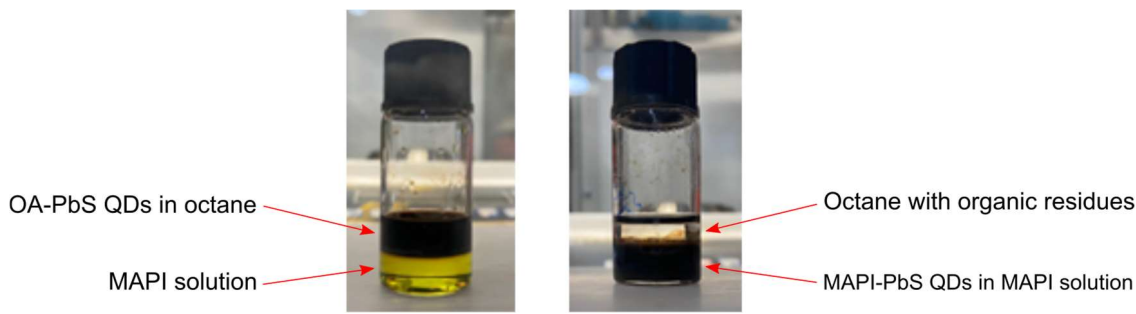

Figure S1. Pictures of CQD solutions before (left) and after (right) the ligand-exchange process.

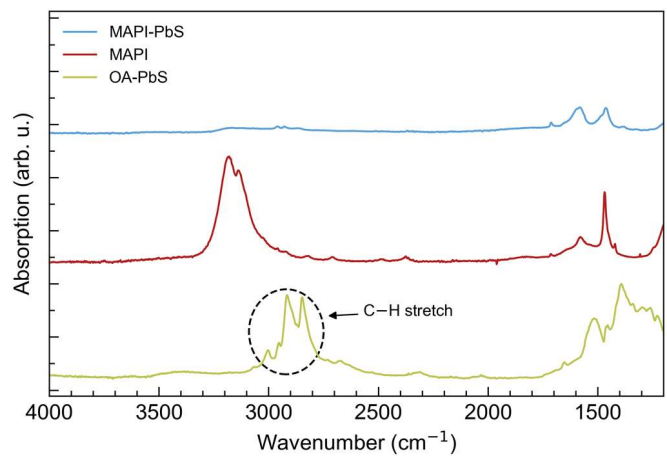

Figure S2. FTIR spectra of PbS QDs before (OA-PbS) and after (MAPI-PbS) the ligand exchange process, together with a reference MAPI film.

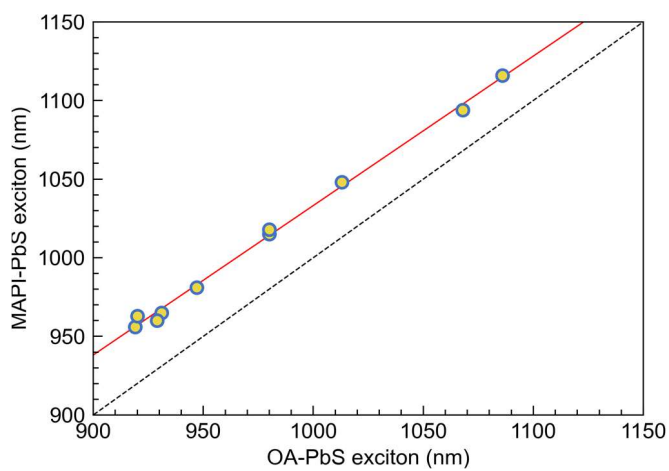

Figure S3. QD excitonic red-shit upon ligand exchange. The red line is a linear regression. The dashed black line marks the case without energy shift.

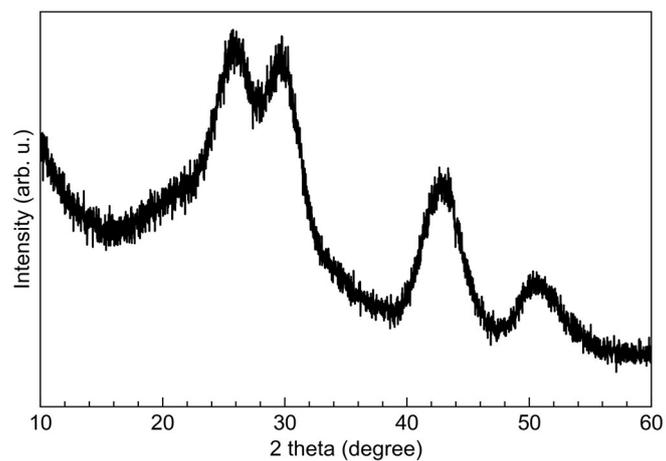

Figure S4. Typical XRD spectrum of OA-PbS QDs.

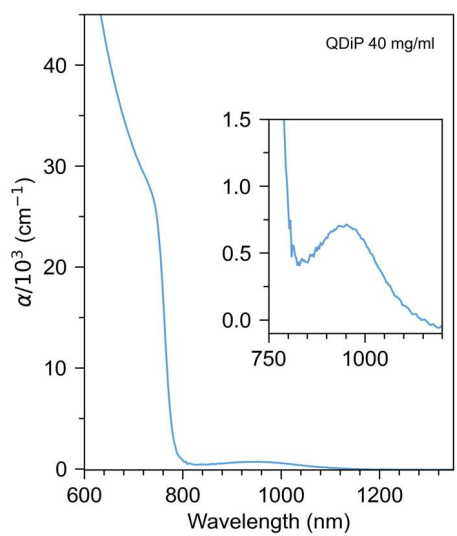

Figure S5. Absorption coefficient of a QDiP material with $C_{\mathrm{QD}}$ = 40 mg/ml.

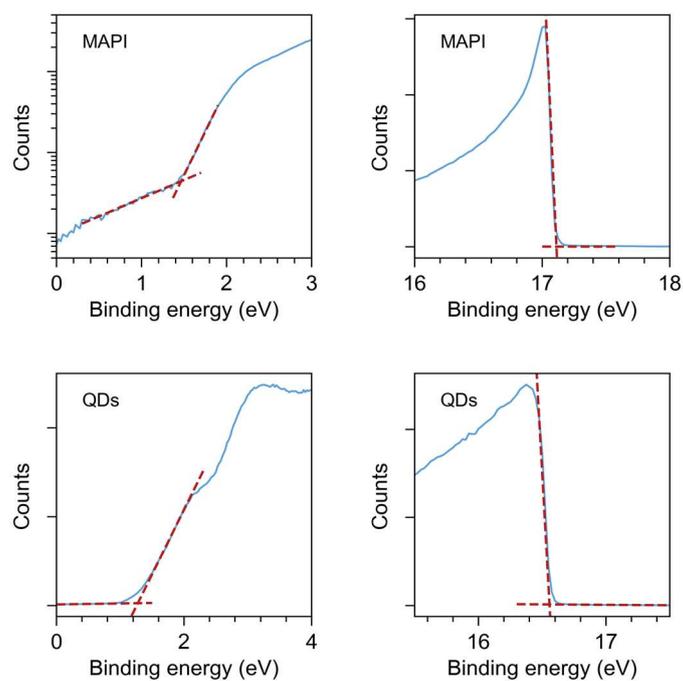

Figure S6. UPS data of a secondary electron cut-off (high-energy fit) and valence band (low-energy fit) of one of the MAPI and one of the MAPI-PbS QDs measured samples.

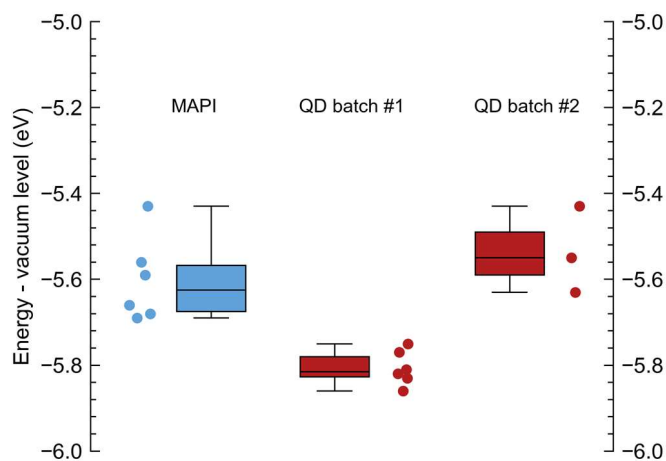

Figure S7. Energies, obtained by fitting of the UPS results, of the VB edge of MAPI and the $1S_h$ state of MAPI-PbS QDs like those used in the measured solar cell. The $1S_e$ energy used in Figure 4a of the main text is the average of the mean values of the two MAPI-QDs batches measured.